\documentclass[12pt]{article}
\usepackage{graphicx, url}
\usepackage{palatino}
\usepackage{amsmath, amssymb,bbm}
\usepackage{epsfig}
\usepackage{rotating}
\usepackage{dcolumn}
\usepackage{bm}
\usepackage{caption}
\usepackage{subcaption}
\usepackage[table]{xcolor}
\newcommand{\comment}[1]{}
\def \non{\nonumber}
\def \ra{\rightarrow}
\def \bea{\begin{eqnarray}}
\def \eea{\end{eqnarray}}

\def \bbbar{b\overline{b}}

\def \sg{\sigma}
\begin{document}   
\baselineskip 18pt
\title{Probing $\Upsilon({1S})$, $\eta_b$ and $h_b$ resonances 
in proton-lead collisions at the LHCb}
\author{
   Sudhansu~S.~Biswal$^1$\footnote{E-mail: ssbiswal@ravenshawuniversity.ac.in}, 
   ~Priyabrata Dash$^1$\footnote{E-mail: priyabrata2023@ravenshawuniversity.ac.in}\\
   ~Sushree~S.~Mishra$^2$\footnote{E-mail: sushreesimran.mishra97@gmail.com},
     ~and  K.~Sridhar$^3$\footnote{E-mail: sridhar.k@apu.edu.in} \\ [0.2cm]
    {\it \small 1. Department of Physics, Ravenshaw University,} \\ [-0.2cm]
    {\it \small Kataka, 753003, India.}\\ [-0.2cm]
        {\it \small 2. Department of Physics, Stewart Science College,} \\ [-0.2cm]
    {\it \small Kataka, 753001, India.}\\ [-0.2cm]
    {\it \small 3. School of Arts and Sciences, Azim Premji University,} \\ [-0.2cm]
    {\it \small Sarjapura, Bangalore, 562125, India.}\\
}
\date{}
\maketitle

\begin{abstract}

We explore the production of bottomonium states in proton--lead (pPb) collisions 
at the Large Hadron Collider (LHC) within the frameworks of 
Non-Relativistic Quantum Chromodynamics (NRQCD) and modified NRQCD (MNRQCD). 
In this work, we present theoretical predictions for the production cross-section, 
the nuclear modification factor $R_{\rm pPb}$ and the forward-to-backward 
production ratio $R_{\rm FB}$ of $\Upsilon(1S)$ as functions of 
transverse momentum and rapidity in the LHCb kinematic region. 
The MNRQCD framework, which incorporates perturbative soft-gluon 
emissions from color-octet states, provides description of 
quarkonium production in heavy-ion collision. 
Furthermore, using heavy-quark spin symmetry, 
we estimate the production cross-sections and expected event yields 
of the spin-singlet bottomonium states $\eta_b$ and $h_b$ 
within both NRQCD and MNRQCD. The predictions show significant differences 
between the two approaches in the integrated cross-sections and 
transverse momentum distributions of these resonances. 
Therefore, these results could provide important benchmarks for 
future experimental measurements at LHCb and contribute to a deeper 
understanding of bottomonia production and nuclear modifications 
in proton--lead collisions.

\end{abstract}

\maketitle

\noindent
Heavy quarkonium \cite{Lan, Brambilla} has become one of the most 
important probes for investigating the properties of strongly 
interacting matter under extreme conditions. Extensive experimental studies 
have been carried out at the Relativistic Heavy Ion Collider 
(RHIC) \cite{PHENIX:2004vcz,STAR:2005gfr} and the Large Hadron Collider 
(LHC) \cite{CMS,LHCbf,ALICEf,ATLASf}, complemented by significant 
theoretical developments in both perturbative and 
non-perturbative Quantum Chromodynamics (QCD). 
In particular, bottomonium states are regarded as especially 
clean probes of the medium because of their large bottom-quark 
mass and relatively small regeneration contribution compared with charmonium.

Unlike proton--proton (pp) collisions, proton--lead (pPb) collisions are 
not expected to produce a long-lived QGP \cite{kha} and therefore provide 
an ideal environment for investigating cold nuclear matter effects. 
These effects, arising from nuclear modifications of the parton distribution 
functions and other initial-state nuclear phenomena, alter heavy-quarkonium 
production \cite{Gavai, Satz, Basu, Zhou, Du, Andronic} relative to pp collisions. 
Their impact is commonly quantified through the nuclear modification factor 
$R_{\rm pPb}$, while the forward-to-backward production ratio $R_{\rm FB}$ 
provides a sensitive probe of the rapidity dependence of nuclear effects. 
Consequently, measurements of bottomonium production in pPb collisions are 
essential for resolving cold nuclear matter (CNM) effects from hot-medium 
effects observed in nucleus--nucleus collisions and for improving our 
understanding of heavy-quarkonium production in a nuclear environment.

Since the bottom quark is significantly heavier than the QCD scale, 
bottomonium production can be treated perturbatively at short distances, 
while its subsequent hadronization into physical bound states is governed 
by long-distance dynamics described within the 
Non-Relativistic Quantum Chromodynamics (NRQCD) \cite{bbl} framework. 
Proton--lead (pPb) collisions offer a unique environment to investigate 
these production mechanisms in the presence of nuclear matter without the 
dominant effects of a hot and deconfined medium. Consequently, studies of 
bottomonium production in pPb collisions are essential for establishing 
the cold nuclear matter baseline, constraining theoretical models and 
providing a reliable reference for interpreting heavy-ion collision measurements.

Bottomonia are among the most reliable probes for investigating heavy-quarkonium 
production in high-energy collisions owing to the large mass of the bottom quark. 
Their production can be systematically described within the framework of NRQCD, 
which exploits the non-relativistic nature of the heavy quark--antiquark $Q\bar{Q}$ 
system, where the relative velocity of the constituents is much smaller than the speed of light.

Within the NRQCD factorization formalism, the heavy $Q\bar{Q}$ pair is 
produced perturbatively at short distances in either a color-singlet or 
a color-octet state. The subsequent evolution into a physical bottomonium 
state is governed by long-distance matrix elements (LDMEs). 
While the color-singlet mechanism contributes at leading order, 
color-octet channels, which are suppressed by powers of the relative velocity (v), 
play a crucial role in accurately describing heavy-quarkonium production 
and are connected to the physical bottomonium state through the emission of soft gluons.

The cross-section for the production of a 
quarkonium state $H$ of mass $M$ in NRQCD can be expressed as:

\bea
  \sigma(H)\;=\;\sum_{n=\{\alpha,S,L,J\}} {F_n\over {M}^{d_n-4}}
       \langle{\cal O}^H_n({}^{2S+1}L_J)\rangle, 
\label{factorizn}
\eea
where $F_n$'s are the short-distance coefficients 
and ${\cal O}_n$ are operators of naive dimension $d_n$, 
describing the long-distance effects. Due to NRQCD factorization, 
the non-perturbative matrix elements are energy independent 
and can be extracted at a given energy and used in the 
prediction of quarkonium cross-sections at other energies.

NRQCD has been more successful in explaining the systematics 
of quarkonium production at the Fermilab Tevatron~\cite{cdf,CDF:2001fdy}, 
compared to the then existing 
Color Singlet Model (CSM) \cite{br}, which was used to analyze 
the production of quarkonia, where the $Q \bar Q$ state produced 
in the short-distance process was assumed to be a color-singlet.
NRQCD predicts transverse polarisation for quarkonia production at high $p_T$, 
but experiments fail to see any evidence for the polarisation 
in $J/\psi$~\cite{jpsi_pol} or $\Upsilon$~\cite{CDF:2001fdy, CMS1, LHCb2} measurements.
Therefore, independent tests of 
NRQCD~\cite{Sridhar:1996vd,Sridhar:2008sc,tests,Mathews:1998nk,bs1,test1} 
are important and the prediction of polarisation of the produced
quarkonium state is an important test to study the discrepancy between the 
predictions made by NRQCD for the quarkonium polarisation at high $p_T$, 
thereby addressing a longstanding puzzle in high-energy collisions.

In Refs.~\cite{bms12, bms3, bms5, Biswal:2023xnk}, 
we have studied quarkonia production, where we get a 
significant difference between the NRQCD and 
modified NRQCD (MNRQCD) model in case of $\eta_c$, $\eta_b$, 
$h_c$ and $h_b$ production. 
However, MNRQCD predictions for $\eta_c$ and $h_c$ production 
show good agreement with LHCb experimental results. 
In this paper, we have focused our analysis on bottomonia production in 
proton-lead collisions using both NRQCD and MNRQCD models.

The NRQCD formula for $\Upsilon(1S)$ can be written explicitly in terms 
of the various octet and singlet intermediate states:
\begin{eqnarray}
\sigma_{\Upsilon(1S)}  = \hat F_{{}^{3}S_1^{[1]}} \times \langle {\cal O} ({}^{3}S_1^{[1]}) \rangle +
                \hat F_{{}^{3}S_1^{[8]}} \times \langle {\cal O} ({}^{3}S_1^{[8]}) \rangle +\cr
                 \hat F_{{}^{1}S_0^{[8]}} \times \langle {\cal O} ({}^{1}S_0^{[8]}) \rangle 
                + {1 \over M^2} \biggl\lbrack\hat F_{{}^{3}P_J^{[8]}} \times \langle {\cal O} 
                     ({}^{3}P_J^{[8]}) \rangle \biggr\rbrack .
\label{Fock}
\end{eqnarray}

The above formula gets modified to the following in the MNRQCD with perturbative soft gluon emission:

\begin{eqnarray}
\sigma_{\Upsilon(1S)} &=& \biggl\lbrack \hat F_{{}^{3}S_1^{[1]}} 
                \times \langle {\cal O} ({}^{3}S_1^{[1]}) \rangle \biggr\rbrack \cr 
                &+& \biggl\lbrack  
                  \hat F_{{}^{3}S_1^{[8]}} 
                 + \hat F_{{}^{1}P_1^{[8]}} 
                + \hat F_{{}^{1}S_0^{[8]}} + (\hat F_{{}^{3}P_J^{[8]}} ) \biggr\rbrack 
                \times ({\langle {\cal O} ({}^{3}S_1^{[1]}) \rangle \over 8}) \cr
                &+& \biggl\lbrack  
                  \hat F_{{}^{3}S_1^{[8]}} 
                 + \hat F_{{}^{1}P_1^{[8]}} 
                + \hat F_{{}^{1}S_0^{[8]}} + (\hat F_{{}^{3}P_J^{[8]}} ) \biggr\rbrack 
                \times \langle {\cal O}  \rangle ,
\end{eqnarray}
where
\begin{equation}
     \langle {\cal O}  \rangle =
                \times \biggl\lbrack 
                 \langle {\cal O} ({}^{3}S_1^{[8]}) \rangle 
                + \langle {\cal O} ({}^{1}S_0^{[8]}) \rangle 
                + {\langle {\cal O} ({}^{3}P_J^{[8]}) \rangle \over M^2}
                    \biggr\rbrack. 
\end{equation}

The differential cross section for $b\bar b$ pair production with specific 
angular momentum and color states at the LHC is given by:

\bea
&&\frac{d\sg}{dp_{_T}} \;(p  ~Pb \ra \bbbar\; [^{2S+1}L^{[1,8]}_J]\, X)= \non \\
&&\sum \int \!dy \int \! dx_1 ~x_1\:G_{a/p} (x_1)~x_2\:G_{b/Pb}(x_2) 
\:\frac{4p_{_T}}{2x_1-\overline{x}_T\:e^y}\non\\
&&\frac{d\hat{\sg}}{d\hat{t}}
(ab\ra \bbbar[^{2S+1}L_J^{[1,8]}]\;d),
\label{eq:diff}
\eea
where the summation is over the partons ($a$ and $b$),    
the final state $b\bar b$ is in the $^1 S^{[8]}_0$, $^1 P^{[8]}_1$,  
$^3 S^{[8]}_1$ states and $G_{a/p}$,   
$G_{b/Pb}$ are the distributions of partons $a$ and 
$b$ in the proton and lead respectively. Here, $x_1$ and $x_2$ are 
the respective  momentum they carry. 
In the above formula, 
$\overline{x}_T=\sqrt{x_T^2+4\tau} \equiv 2 M_T/\sqrt{s}$ \ with 
\  $x_T=2p_{_T}/\sqrt{s}$ and  \(\tau=M^2/s\).
$\sqrt{s}$ is the center-of-mass energy, $M$ is the mass of the resonance 
and $y$ is the rapidity at which the resonance is produced. 
The non-perturbative parameters used in our analysis are taken from
Refs. \cite{Braaten:2000cm, Feng:2015wka, Han:2014kxa, Li:2019anc} and 
the bottom-quark mass is 4.18 GeV \cite{ParticleDataGroup:2024cfk}.
Here nuclear Parton Distribution Functions \cite{Kovarik} for the nucleus are as follows:
\bea
 G_{b/Pb}(x_{2}) = \dfrac{Z}{A}G^{p/A}(x_{2})+\dfrac{A-Z}{A}G^{n/A}(x_{2}).
\eea
The fixed-order perturbative calculations
have been used to get the cross-section for bottomonia production 
and a cut-off is imposed in the calculations for
low-$p_T$ regime. The bottomonia cross-section in the low-$p_T$ 
region requires a resummation of multiple gluon radiation.

Our analysis tests the energy independence of the extracted non-perturbative parameter. 
Unless given by some factorisation theorem, 
this independence cannot be assumed a priori. The agreement with LHC data demonstrates 
that the non-perturbative parameter remains energy independent, as theoretically expected.

\begin{figure}[!h]
\begin{center}
\includegraphics[width=15cm, height=9cm]{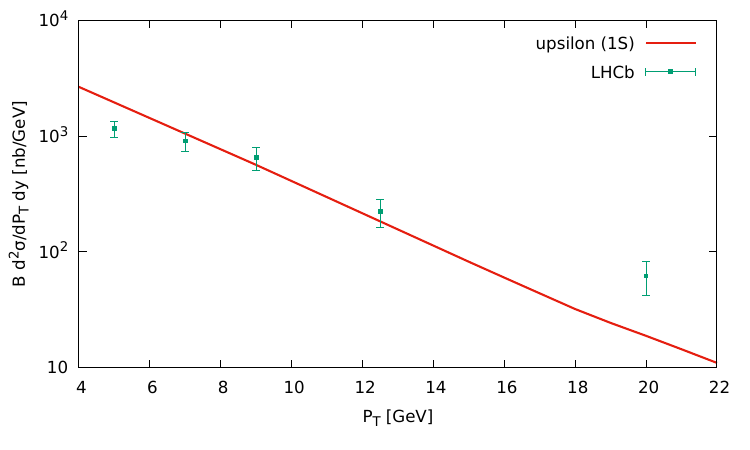}
\caption{Predicted differential distributions for $\Upsilon(1S)$ production
in pPb collision compared to the data from the LHCb experiment.}
        \label{fig:fig1}
\end{center}
\end{figure}

\begin{figure}[!h]
\begin{center}
\includegraphics[width=15.5cm, height=8cm]{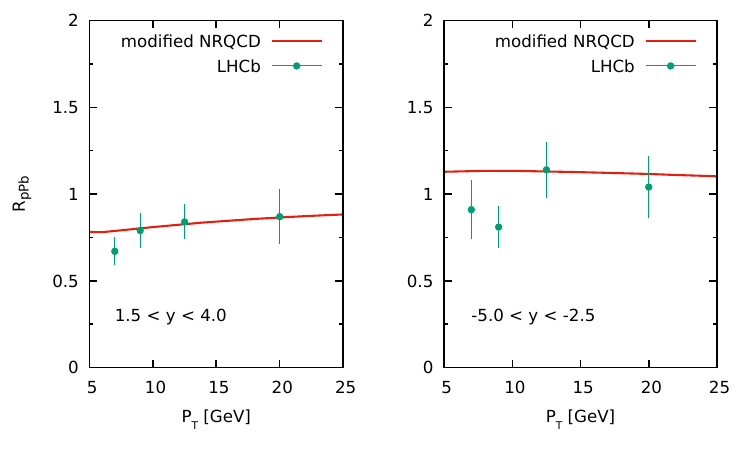}
\caption{Transverse momentum dependence of $R_{\rm pPb}$ for 
$\Upsilon(1S)$ production in pPb collisions at the LHC.}
        \label{fig:fig2}
\end{center}
\end{figure}

\begin{figure}[!h]
\begin{center}
\includegraphics[width=15.5cm, height=8cm]{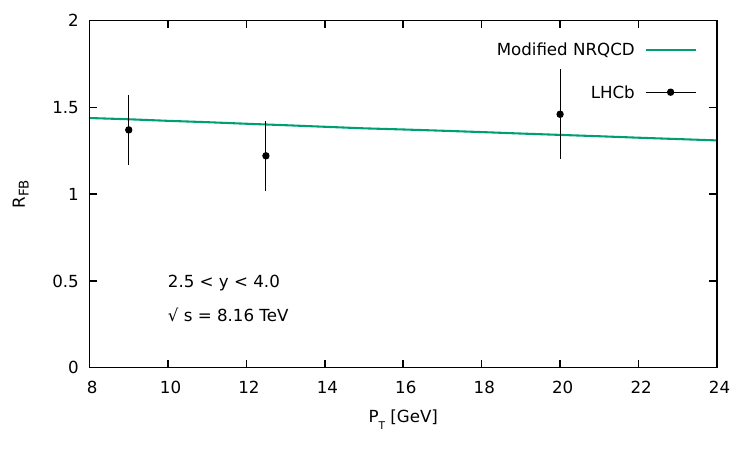}
\caption{Transverse momentum 
dependence of $R_{\rm FB}$ for $\Upsilon(1S)$ production 
in pPb collisions at the LHC.}
        \label{fig:fig3}
\end{center}
\end{figure}

In Fig. \ref{fig:fig1}, the theoretical production cross-sections for $\Upsilon(1S)$ 
have been compared with the LHCb \cite{LHCb:2018psc}  experimental data for 
prompt production, showing an agreement with MNRQCD for pPb collisions. 
Here, we have used the fitted parameter for $\Upsilon(1S)$ production 
to be -1.31 GeV$^3$ \cite{Biswal:2023xnk}.

Nuclear modification factor is defined as the ratio of the yield in heavy-ion collisions 
to that in pp collisions, scaled by the average number of binary collisions. 
Here the nuclear modification factor for pPb configuration is:

\begin{equation}
      R_{pPb}(p_T, |y|) = \frac{1}{208} \frac{d^2\sigma_{pPb}/dp_{T}dy}{d^2\sigma_{pp}/dp_{T}dy},
\end{equation}

where, $\sigma_{pp}$ is the cross-section from pp collisions.

The formula for forward-to-backward ratio is defined as:

\begin{equation}
      R_{FB}(p_T, |y|) = \frac{d^2\sigma_{pPb}(p_T, +|y|)/dp_{T}dy}{d^2\sigma_{pPb}(p_T, -|y|)/dp_{T}dy}.
\end{equation}

Figs. \ref{fig:fig2} and \ref{fig:fig3} show the nuclear modification factor 
and forward-to-backward production ratio predictions using MNRQCD, 
which gives significant results while comparing with the experimental data. 
Here, the MNRQCD successfully reproduces the kinematic dependence of both the 
nuclear modification factor ($R_{\rm pPb}$) and the forward-to-backward ratio 
($R_{\rm FB}$) across the transverse momentum and rapidity ranges, 
while comparing with the experimental data.
The Figs. \ref{fig:fig2} and \ref{fig:fig3} suggest
that $R_{\rm pPb}$ and $R_{\rm FB}$ work as a good observable up to at least 
$p_T \sim$ 20~GeV, exhibiting a stable behavior around unity.

In Figs. \ref{fig:fig2} and \ref{fig:fig3}, the LHC data, we have taken in the 
different rapidity ranges 
have relatively larger statistical and systematic uncertainties. 
Hence, the theoretical predictions based on these bin-averaged values 
may not fully reproduce the detailed shape of the experimental distributions.

\begin{figure}[h]
\begin{center}
\includegraphics[width=12.5cm, height=7.2cm]{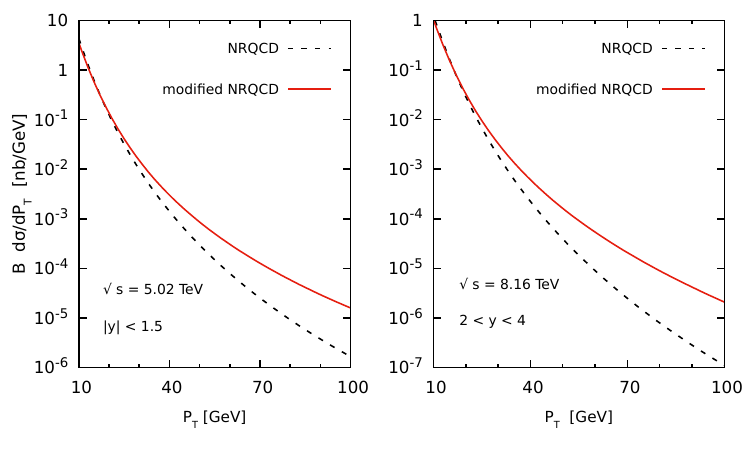}
\caption{ Predicted differential cross-sections for $\eta_{b}$ production 
in pPb collisions at the LHC. }
        \label{fig:fig4}
\end{center}
\end{figure}

\begin{figure}[!h]
\begin{center}
\includegraphics[width=12.5cm, height=7.2cm]{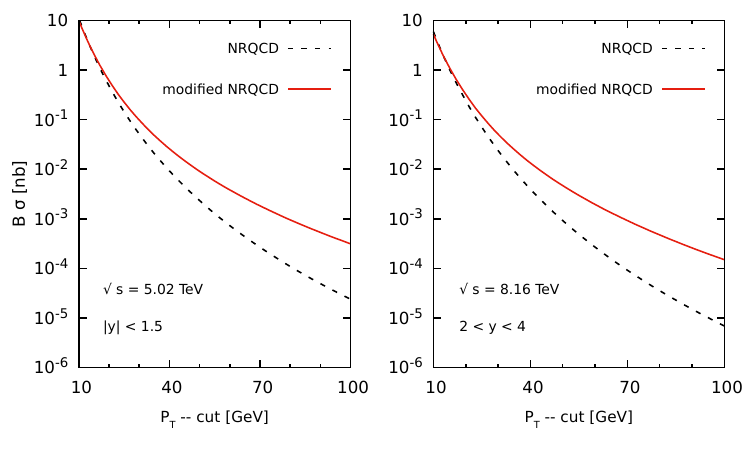}
\caption{ Predicted Integrated cross-sections for $\eta_{b}$ production 
in pPb collisions at the LHC. }
        \label{fig:fig5}
\end{center}
\end{figure}

\begin{table}[htbp]
\centering
 \hspace*{-0.1cm}
\begin{tabular}{ |m{0.5cm}|m{0.5cm}|m{0.5cm}|m{0.5cm}|m{0.5cm}| }
 \hline
\multicolumn{1}{|c|}{ } &\multicolumn{4}{c|}{$\sim$Expected number of events} \\[1mm]

 \cline{2-5} 
\multicolumn{1}{|c|}{} &\multicolumn{2}{c|}{}
&\multicolumn{2}{c|}{} \\[-2mm]

\multicolumn{1}{|c|}{} &\multicolumn{2}{c|}{Rapidity: $|y| < 1.5$, $\sqrt{s}$ = 5.02 TeV}
&\multicolumn{2}{c|}{Rapidity: $2 < y < 4$, $\sqrt{s}$ = 8.16 TeV} \\[1mm]
\cline{2-5}
&&&& \\[-3mm]
\multicolumn{1}{|c|}{Model} 
	&\multicolumn{1}{c|}{$P_{T}$ $>$ 5 GeV}&\multicolumn{1}{c|}{$P_{T}$ $>$ 10 GeV}
&\multicolumn{1}{c|}{$P_{T}$ $>$ 5 GeV}&\multicolumn{1}{c|}{$P_{T}$ $>$ 10 GeV}
 \\[1mm]

\hline
\hline

\multicolumn{1}{|c|}{NRQCD} &
\multicolumn{1}{c|}{$8.2 \times 10^{5}$} &\multicolumn{1}{c|}{$1.1 \times 10^{5}$}
&\multicolumn{ 1}{c|}{$4.6 \times 10^{5}$}&\multicolumn{1}{c|}{$5.9 \times 10^{4}$}  
 \\
\hline
\multicolumn{1}{|c|}{MNRQCD} &
\multicolumn{1}{c|}{$6.4 \times 10^{5}$} &\multicolumn{1}{c|}{$9.9 \times 10^{4}$}
&\multicolumn{1}{c|}{$3.6 \times 10^{5}$} &\multicolumn{1}{c|}{$5.2 \times 10^{4}$}   
\\
 \hline 
  
\end{tabular}
\caption{\label{tab:events1}
Number of $\eta_b$ events expected at the LHC correspond to 
integrated luminosity of 100 nb$^{-1}$. 
}
\end{table}

\begin{figure}[h!]
\begin{center}
\includegraphics[width=12.5cm, height=7.7cm]{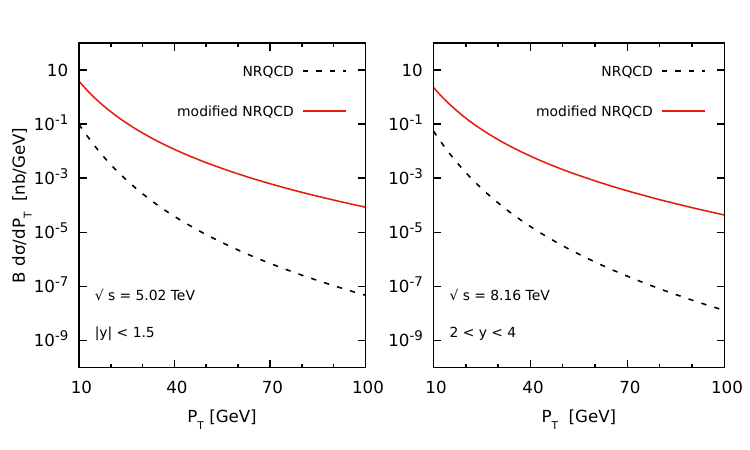}
\caption{ Predicted differential cross-sections for $h_{b}$ production 
in pPb collisions at the LHC.}
        \label{fig:fig6}
\end{center}
\end{figure}

\begin{figure}[h!]
\begin{center}
\includegraphics[width=12.5cm, height=7.7cm]{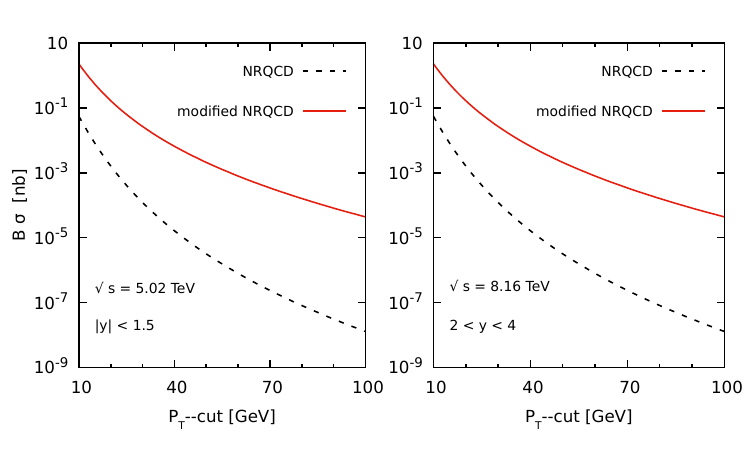}
\caption{ Predicted Integrated cross-sections for $h_{b}$ production 
in pPb collisions at the LHC.}
        \label{fig:fig7}
\end{center}
\end{figure}

For better understanding, we have extended our work on 
$p_T$ distributions of $\eta_b$ and $h_b$ production 
in both NRQCD and MNRQCD.
Figs. \ref{fig:fig4} and \ref{fig:fig5} represent the $\eta_b$ 
production differential cross-sections as a function
of $p_T$ and integrated cross-sections for different $p_T$ - cuts in both NRQCD and
MNRQCD. Similarly, Figs. \ref{fig:fig6} and \ref{fig:fig7} are for $h_b$ production.

\begin{table}[htbp]
\centering
 \hspace*{-0.1cm}
\begin{tabular}{ |p{1cm}|p{1cm}|p{1cm}|p{1cm}|p{1cm}| }
 \hline
\multicolumn{1}{|c|}{ } &\multicolumn{4}{c|}{$\sim$Expected number of events} \\[2mm]

 \cline{2-5} 
\multicolumn{1}{|c|}{} &\multicolumn{2}{c|}{}
&\multicolumn{2}{c|}{} \\[-2mm]

\multicolumn{1}{|c|}{} &\multicolumn{2}{c|}{Rapidity: $|y| < 1.5$, $\sqrt{s}$ = 5.02 TeV}
&\multicolumn{2}{c|}{Rapidity: $2 < y < 4$, $\sqrt{s}$ = 8.16 TeV} \\[2mm]
\cline{2-5}
&&&& \\[-3mm]
\multicolumn{1}{|c|}{Model} 
	&\multicolumn{1}{c|}{$P_{T}$ $>$ 5 GeV}&\multicolumn{1}{c|}{$P_{T}$ $>$ 10 GeV}
&\multicolumn{1}{c|}{$P_{T}$ $>$ 5 GeV}&\multicolumn{1}{c|}{$P_{T}$ $>$ 10 GeV}
 \\[1mm]

\hline
\hline

\multicolumn{1}{|c|}{NRQCD} &
\multicolumn{1}{c|}{$1.8 \times 10^{2}$} &\multicolumn{1}{c|}{$26$}
&\multicolumn{ 1}{c|}{$99$}&\multicolumn{1}{c|}{$13$}   
 \\

\hline

\multicolumn{1}{|c|}{MNRQCD} &
\multicolumn{1}{c|}{$6.2 \times 10^{3}$} &\multicolumn{1}{c|}{$1.4 \times 10^{3}$}
&\multicolumn{1}{c|}{$3.4 \times 10^{3}$} &\multicolumn{1}{c|}{$8.1 \times 10^{2}$}   
\\
 \hline   
\end{tabular}
\caption{\label{tab:events2}
Number of $h_b$ events expected at the LHC correspond to integrated luminosity of 100 nb$^{-1}$ . 
}
\end{table}

In this study, the $h_b$ state is considered through its radiative 
decay ($h_b \rightarrow \eta_b + \gamma$), 
which has a branching fraction of approximately $52\%$. The subsequent decay of 
the ($\eta_b$) state into a proton--antiproton $(p\bar{p})$ pair is assumed 
to have a branching fraction of 
$1.33 \times 10^{-3}$ \footnote{The branching ratio for $\eta_c$ $\rightarrow$ $p\bar{p}$ 
is used, since the branching ratio for $\eta_b$ $\rightarrow$ $p\bar{p}$ is not available}.

To get a sense of the feasibility of measuring the $\eta_b$ and $h_b$ production at the LHC, 
we have calculated the $p_T$ -integrated cross sections in two different rapidity ranges. 
These results, presented in Tables 1 and 2 for an integrated luminosity of 100 nb$^{-1}$, 
suggest that a significant number of $\eta_b$ and $h_b$ events can be expected at the LHC.

The $h_b$ state has been studied through its radiative 
decay $h_b \rightarrow \eta_b + \gamma$, followed by the decay $\eta_b \rightarrow p\bar{p}$, 
using the corresponding branching fractions. Although the experimental observation of 
these states remains challenging because of their small production rates 
and difficult decay channels, the high-luminosity phase of the LHC and future detector 
upgrades are expected to substantially improve their detection prospects. 
Such measurements will provide a stringent test of NRQCD and MNRQCD predictions 
and offer valuable insight into the production mechanisms of 
bottomonium in a nuclear environment. In particular, a comparison of future 
experimental data with our predictions for $\Upsilon(1S)$, $\eta_b$, and $h_b$ 
production will help constrain the non-perturbative inputs of the MNRQCD framework 
and further our understanding of heavy-quarkonium production in proton--lead collisions.

In conclusion, we have investigated the production of $\Upsilon(1S)$ in 
proton--lead (pPb) collisions within the framework of modified 
Non-Relativistic Quantum Chromodynamics (MNRQCD). The production cross-sections, 
nuclear modification factor $R_{\rm pPb}$, and forward-to-backward production 
ratio $R_{\rm FB}$ have been evaluated as functions of transverse momentum and 
rapidity in the LHCb kinematic region. Furthermore, we have presented theoretical 
predictions for the production of the spin-singlet bottomonium states $\eta_b$ 
and $h_b$ within both NRQCD and MNRQCD, highlighting significant differences 
in the integrated cross-sections and transverse momentum distributions 
obtained in the two approaches.



\end{document}